\documentclass{camera}
\usepackage{graphicx}  
\usepackage{hyperref}

\begin{document}

%
\title{The highest energy neutrinos: first evidence for cosmic origin}

%
\author{Francis Halzen}

%
\organization{Wisconsin IceCube Particle Astrophysics Center and Department of Physics, University of Wisconsin, Madison, USA}

\maketitle

\begin{abstract}
Developments in neutrino astronomy have been to a great extent motivated by the search for the sources of the cosmic rays, leading at a very early stage to the concept of a cubic kilometer neutrino detector. Almost four decades later such an instrument, IceCube, is taking data and has produced the first evidence for a flux of high-energy neutrinos of cosmic origin. After a brief review of the history of the field, we will introduce IceCube and describe the first analysis of data taken with the completed instrument. The atmospheric neutrino flux cannot accommodate an excess of 28 neutrinos observed with energies above 60 TeV. We will briefly speculate on the origin of these events. Readers interested specifically in IceCube results may refer directly to section 3.
\end{abstract}

%

\section{A Brief History of Neutrino Astronomy}

Soon after the 1956 observation of the neutrino \cite{Reines1956}, the idea emerged that it represented the ideal astronomical messenger \cite{Greisen1960,Reines1960,Markov1960}. The concept has since been demonstrated: neutrino detectors have ``seen" the Sun and detected a supernova in the Large Magellanic Cloud in 1987. Both observations were of tremendous importance; the former showed that neutrinos have a tiny mass, opening the first chink in the armor of the Standard Model of particle physics, and the latter confirmed the basic nuclear physics of the death of stars.
	
High-energy neutrinos have a distinct potential to probe the extreme Universe. Neutrinos reach us from the edge of the Universe without absorption and with no deflection by magnetic fields. They can escape unscathed from the inner neighborhood of black holes and from the accelerators where cosmic rays are born. Their weak interactions also make neutrinos very difficult to detect. Immense particle detectors are required to collect cosmic neutrinos in statistically significant numbers \cite{Gaisser1995}. Already by the 1970s, it had been understood \cite{Roberts1992} that a kilometer-scale detector was needed to observe the ``cosmogenic" neutrinos produced in the interactions of cosmic rays with background microwave photons \cite{BZ}.

Above a threshold of $\sim 4\times10^{19}$\,eV, cosmic rays interact with the microwave background introducing an absorption feature in the cosmic-ray flux, the Greisen-Zatsepin-Kuzmin (GZK) cutoff. The mean free path of extragalactic cosmic rays propagating in the microwave background is limited to less than 100\, megaparsecs. Therefore, secondary neutrinos produced in these interactions are the only probe of the still enigmatic sources at further distances. Realistic calculations \cite{Stecker} of the neutrino flux associated with the observed flux of extragalactic cosmic rays appeared in the 1970s and predicted on the order of one event per year in a kilometer-scale detector, subject to astrophysical uncertainties. Today's estimates of the sensitivity for observing potential cosmic accelerators such as Galactic supernova remnants, active galactic nuclei (AGN), and gamma-ray bursts (GRB) unfortunately point to the same exigent requirement \cite{Gaisser1995}.  Building a neutrino telescope has been a daunting technical challenge.

Given the detector's required size, early efforts concentrated on instrumenting large volumes of natural water with photomultipliers that detect the Cherenkov light emitted by the secondary particles produced when neutrinos interact with nuclei inside or near the detector \cite{Markov19602}. After a two-decade-long effort, building the Deep Underwater Muon and Neutrino Detector (DUMAND) in the sea off the main island of Hawaii unfortunately failed \cite{Babson1990}. However, DUMAND pioneered many of the detector technologies in use today and inspired the deployment of a smaller instrument in Lake Baikal \cite{Balkanov2003} as well as efforts to commission neutrino telescopes in the Mediterranean \cite{Aggouras2005,Aguilar2006,Migneco2008}.  These have paved the way toward the planned construction of KM3NeT \cite{Migneco2008}. 

The first telescope on the scale envisaged by the DUMAND collaboration was realized instead by transforming a large volume of deep Antarctic ice into a particle detector, the Antarctic Muon and Neutrino Detector Array (AMANDA). In operation from 2000 to 2009, it represented the proof of concept for the kilometer-scale neutrino observatory, IceCube \cite{ICPDD2001,Ahrens2004}, completed in 2010. We present this talk at the critical time that IceCube data taken with the completed detector have revealed the first evidence for a flux of high-energy neutrinos reaching us from beyond the Sun.

\begin{figure*}[t]
\begin{center}
\vspace{-25pt}
\includegraphics[width=0.75\textwidth,trim=0px 30px 50px 20px,clip=true]{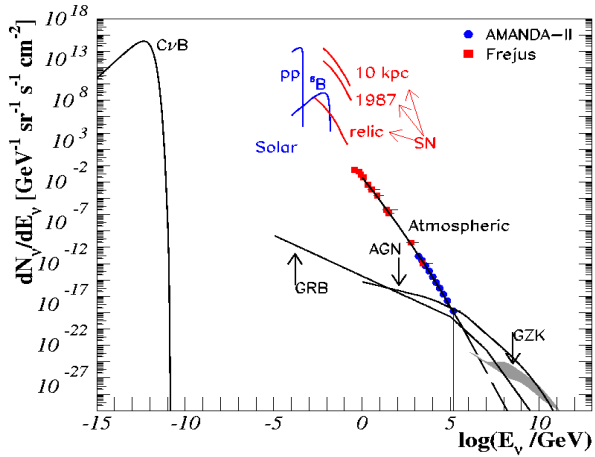}
\caption{The cosmic-neutrino spectrum. Sources are the Big Bang (C$\nu$B), the Sun,
supernovae (SN), atmospheric neutrinos, gamma-ray bursts (GRB), active galactic nuclei (AGN),
and cosmogenic (GZK) neutrinos. The data points are from a detector at the Fr\'{e}jus underground
laboratory \cite{Rhode1996} (red) and from AMANDA \cite{Achterberg2007} (blue). Figure courtesy of J. Becker \cite{Gaisser1995}.}
\label{nu_spectrum}
\end{center}
\end{figure*}

Fig. \ref{nu_spectrum} illustrates the cosmic neutrino energy spectrum covering an enormous range, from the neutrinos produced in association with the 2.725 K microwave photon background to $10^{20}$\,eV \cite{Becker2008}. The figure is a mixture of observations and theoretical predictions. At low energy, the neutrino sky is dominated by neutrinos produced in the Big Bang. At MeV energy, neutrinos are produced by the Sun and by supernova explosions; the flux from the 1987 event is shown. At higher energies, the neutrino sky is dominated by neutrinos produced in cosmic-ray interactions in the atmosphere, measured up to energies of 100\ TeV by the AMANDA experiment \cite{Achterberg2007}. Atmospheric neutrinos are the dominant background when searching for extraterrestrial sources of neutrinos. The flux of atmospheric neutrinos fortunately falls dramatically with increasing energy; events above 100 TeV are rare, leaving a clear field of view of the sky for extraterrestrial sources. In Fig. \ref{nu_spectrum} the cosmogenic flux, previously introduced, shares the high-energy neutrino sky with neutrinos anticipated from gamma-ray bursts and active galactic nuclei \cite{Gaisser1995}.

\section{Rationale for the Construction of Kilometer-Scale Neutrino Detectors}

Despite their discovery potential touching a wide range of scientific issues, the construction of kilometer-scale neutrino detectors has been largely motivated by the prospect of detecting neutrinos associated with cosmic rays.

Cosmic accelerators produce particles with energies in excess of $100$\,EeV; we still do not know where or how \cite{Sommers2009}. The bulk of the cosmic rays are Galactic in origin. Any association with our Galaxy presumably disappears at EeV energy when the gyroradius of a proton in the Galactic magnetic field exceeds its size. The cosmic-ray spectrum exhibits a rich structure above an energy of $\sim 0.1$\,EeV, but where exactly the transition to extragalactic cosmic rays occurs is a matter of debate.

The detailed blueprint for a cosmic-ray accelerator must meet two challenges: the highest-energy particles in the beam must reach beyond $10^3$\,TeV ($10^8$\,TeV) for Galactic (extragalactic) sources, and their luminosities must be able to accommodate the observed flux. Both requirements represent severe constraints that have limited theoretical speculations.

Supernova remnants were proposed as possible sources of Galactic cosmic rays as early as 1934 by Baade and Zwicky \cite{Baade1934}; their proposal is still a matter of debate after more than 75 years \cite{Butt2009}. The idea is generally accepted because of energetics: three Galactic supernova explosions per century converting a reasonable fraction of a solar mass into particle acceleration can accommodate the steady flux of cosmic rays in the Galaxy. Energetics also drives speculations on the origin of extragalactic cosmic rays.

By integrating the cosmic-ray spectrum above the ankle at $\sim4$\,EeV, we find that the energy density of the Universe in extragalactic cosmic rays is $\sim 3 \times 10^{-19}\rm\,erg\ cm^{-3}$ \cite{Gaisser1997}.  This value is rather uncertain because of our ignorance of the precise energy where the transition from Galactic to extragalactic sources occurs. The power required for a population of sources to generate this energy density over the Hubble time of $10^{10}$\,years is $2 \times 10^{37}\,\rm\,erg\ s^{-1}$ per Mpc$^3$. A gamma-ray-burst fireball converts a fraction of a solar mass into the acceleration of electrons, seen as synchrotron photons. The observed energy in extragalactic cosmic rays can be accommodated with the reasonable assumption that shocks in the expanding GRB fireball convert roughly equal energy into the acceleration of electrons and cosmic rays \cite{Waxman1995}. It so happens that $2 \times 10^{51}$\,erg per GRB will yield the observed energy density in cosmic rays after $10^{10}$ years, given that their rate is on the order of 300 per $\textrm{Gpc}^{3}$ per year. Hundreds of bursts per year over Hubble time produce the observed cosmic-ray density, just like three supernovae per century accommodate the steady flux in the Galaxy.

Problem solved? Not really: it turns out that the same result can be achieved assuming that active galactic nuclei convert, on average, $2 \times 10^{44}\,\rm\,erg\ s^{-1}$ each into particle acceleration \cite{Gaisser1995}. As is the case for GRBs, this is an amount that matches their output in electromagnetic radiation.  Whether GRBs or AGN, the observation that cosmic-ray accelerators radiate similar energies in photons and cosmic rays may not be an accident.

Neutrinos are produced in association with the cosmic-ray beam.  Cosmic rays accelerated in regions of high magnetic fields near black holes or neutron stars inevitably interact with radiation surrounding them. Cosmic-ray accelerators are beam dumps. In supernova shocks, cosmic rays inevitably interact with the hydrogen in the Galactic disk, producing equal numbers of pions of all three charges that decay into pionic photons and neutrinos. Their secondary fluxes should be boosted by the interaction of the cosmic rays with high-density molecular clouds that are ubiquitous in the star-forming regions where supernovae are more likely to explode. For extragalactic sources, the neutrino-producing target may be light, for instance photons radiated by the accretion disk of an AGN, or synchrotron photons that coexist with protons in the expanding fireball producing a GRB.

\begin{figure*}[t]
\begin{center}
\vspace{-25pt}
\includegraphics[width=1.0\textwidth,trim=0px 70px 0px 10px,clip=true]{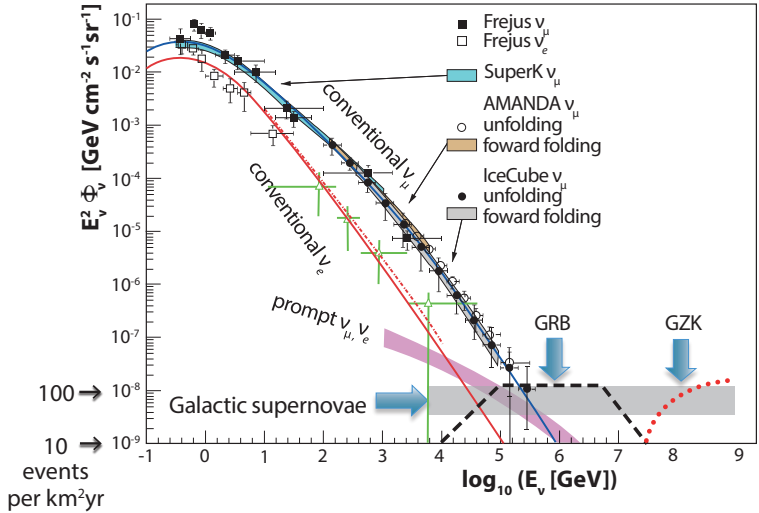}
\caption{Anticipated cosmic-neutrino fluxes produced by supernova remnants and GRBs exceed the atmospheric neutrino flux in IceCube above 100\, TeV. Also shown is a sample calculation of the cosmogenic neutrino flux. The atmospheric electron-neutrino spectrum (green open triangles) is from \cite{Aartsen:2013rt}. The conventional $\nu_e$ (red line) and $\nu_\mu$ (blue line) from Honda, $\nu_e$ (red dotted line) from Bartol and charm-induced neutrinos (magenta band) \cite{Enberg:2008te} are shown. Previous measurements from Super-K \cite{GonzalezGarcia:2006ay}, Frejus \cite{Frejus}, AMANDA \cite{abbasi2009,abbasi2010} and IceCube \cite{abbasi2011a,abbasi2011} are also shown.}
\label{discovery}
\end{center}
\end{figure*}

Estimating the neutrino flux associated with cosmic rays accelerated in supernova remnants and GRBs is relatively straightforward as both the beam, identified with the observed cosmic-ray flux, and the targets, observed by astronomers, are known. As was the case for cosmogenic neutrinos, the results, shown in Fig. \ref{discovery}, are subject to astrophysical uncertainties. However, the message is clear, neutrinos from theorized cosmic-ray accelerators dominate the steeply falling atmospheric neutrino flux above an energy of $\sim100$\,TeV. The level of events observed in a cubic-kilometer neutrino detector is $10\sim100$ per year. These estimates reinforced the logic for building a cubic kilometer neutrino detector. A more detailed description of the theoretical estimates can be found in reference \cite{Halzen:2013bta}.

\begin{figure*}[htb]
\begin{center}
\includegraphics[width=0.8\textwidth]{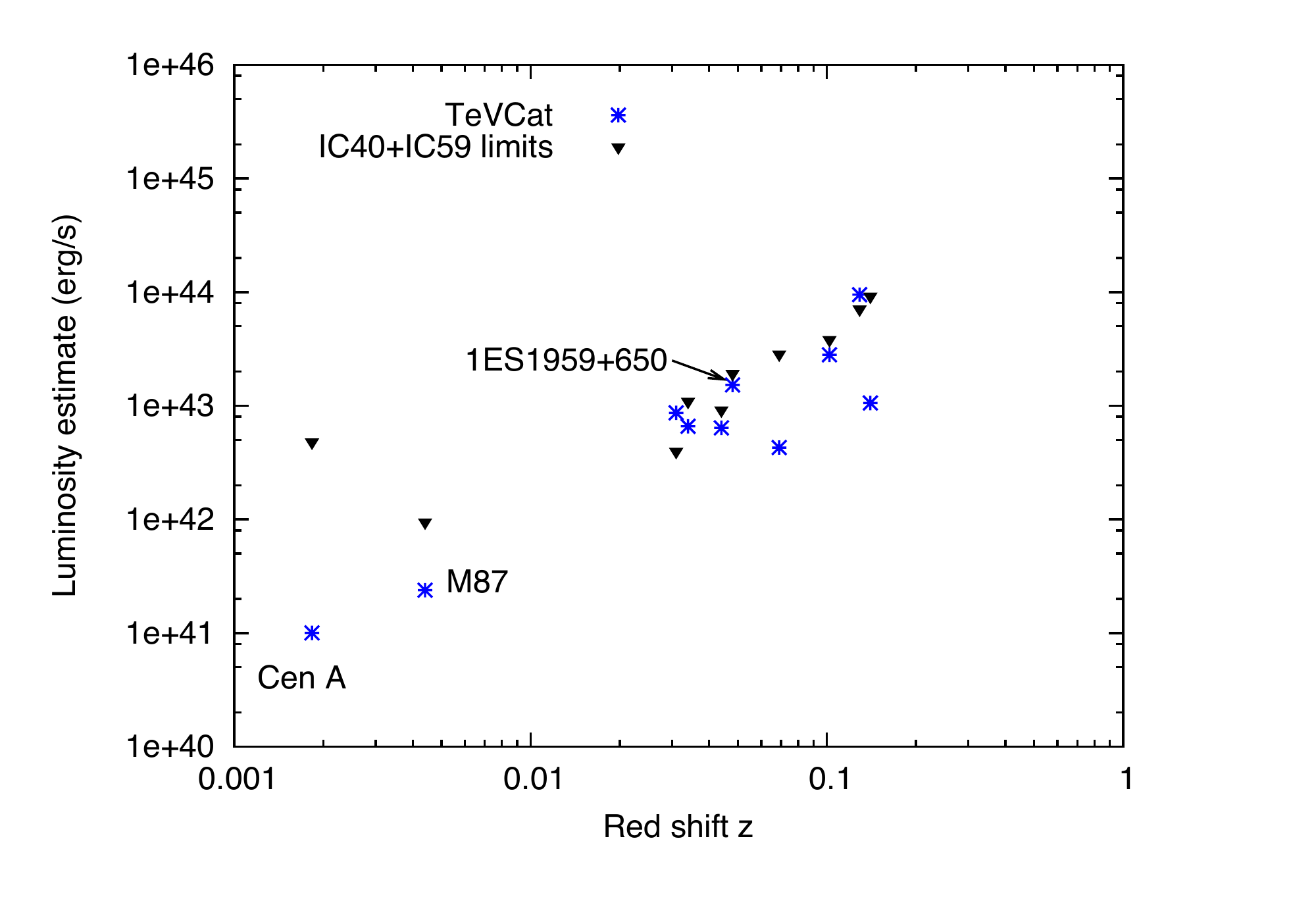}
\caption{Limits on the neutrino flux from selected active galaxies derived from IceCube data taken during construction when the instrument was operating with 40 and 59 strings of the total 86 instrumented strings of DOMs. These are compared with the TeV photon flux for nearby AGN. Note that energy units are in ergs, not TeV.}
\label{agnem}
\end{center}
\end{figure*}

AGN are complex systems with many possible sites for accelerating cosmic rays and for targets to produce neutrinos. No generic prediction of the neutrino flux exists. However, we have introduced the rationale that generic cosmic-ray sources produce a neutrino flux comparable to their flux of cosmic rays \cite{Gaisser1997} and pionic TeV gamma rays \cite{Alvarez-Muniz2002}. In this context, we introduce Fig.~\ref{agnem} showing the present IceCube upper limits on the neutrino flux from nearby AGN as a function of their distance. Also shown is the TeV gamma-ray flux from the same sources. Except for CenA and M87, the muon-neutrino limits have reached the level of the TeV photon flux. One can sum the sources shown in the figure into a diffuse flux. The result, after dividing by $4\pi/c$ to convert the point source to a diffuse flux, is $3 \times 10^{-12}\,\rm TeV\,cm^{-2}\,s^{-1}\,sr^{-1}$, or approximately $10^{-11}\,\rm TeV\,cm^{-2}\,s^{-1}\,sr^{-1}$ for all neutrino flavors. This is at the level of the generic cosmic-neutrino flux argued for in Fig.~\ref{discovery}.

\subsection{The First Kilometer-Scale Neutrino Detector: IceCube}

\begin{figure*}[htp] 
\begin{center}
\includegraphics[width=0.5\textwidth]{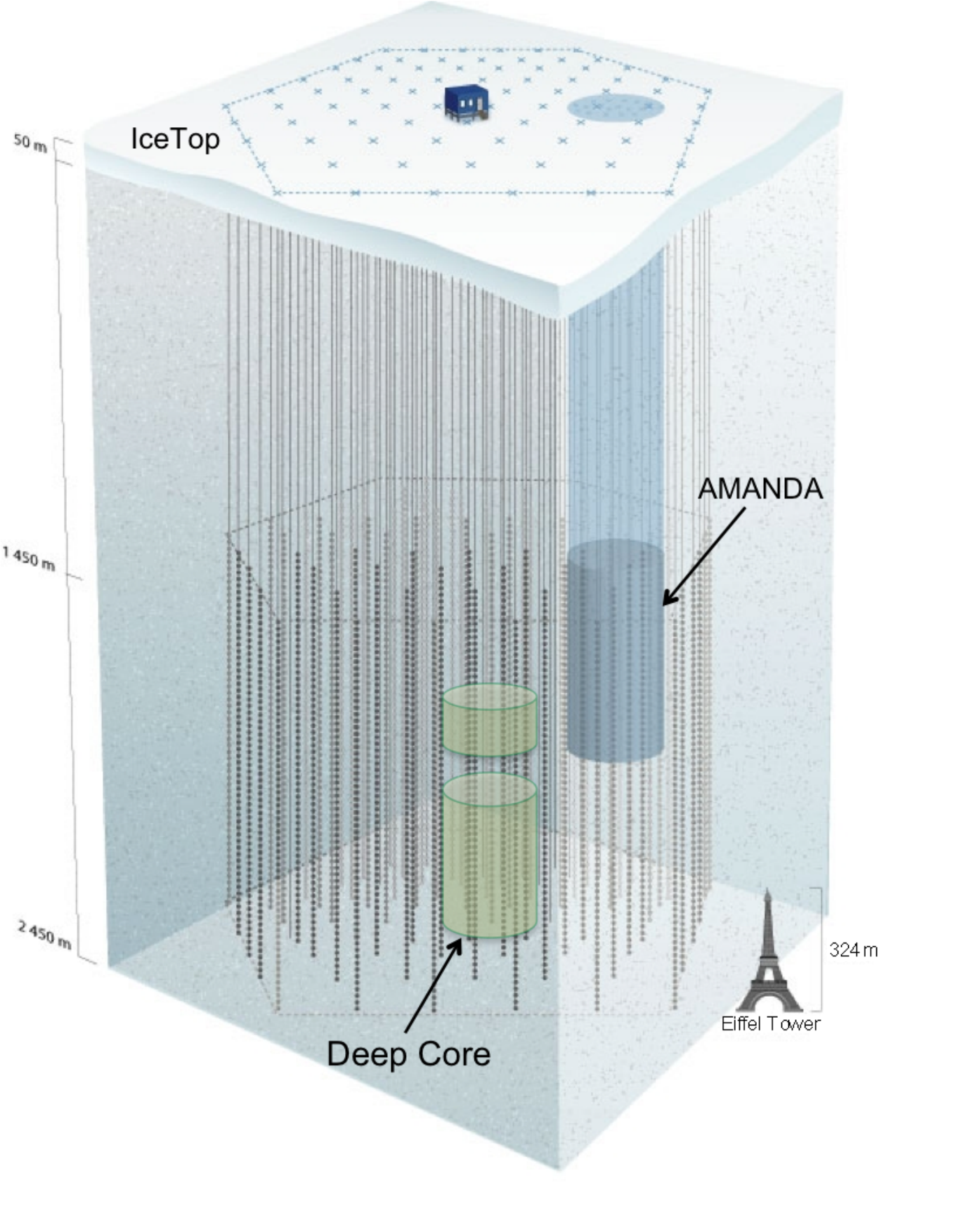}
\caption{Schematic of the IceCube detector.}
\label{deepcore}
\end{center}
\end{figure*}

A series of first-generation experiments \cite{Spiering2009,Katz2011} have demonstrated that high-energy neutrinos with $\sim$\,10\,GeV energy and above can be detected 
using large volumes of highly transparent ice or water instrumented with a lattice of photomultiplier tubes. Such instruments detect neutrinos by observing Cherenkov radiation from secondary particles produced in neutrino interactions inside the detector. Construction of the first second-generation detector, IceCube, at the geographic South Pole was completed in December 2010 \cite{Klein2010}; see Fig. \ref{deepcore}.

IceCube consists of 86 strings, each instrumented with 60 ten-inch photomultipliers spaced 17\,m apart over a total length of one kilometer. The 
deepest modules are located at a depth of 2.45\,km so that the instrument is shielded from the large background of cosmic rays at the surface by approximately 1.5\,km
of ice. Strings are arranged at apexes of equilateral triangles that are 125\,m on a side. The instrumented detector volume is a cubic kilometer of dark and highly transparent \cite{Aartsen:2013rt} Antarctic ice. The radioactive background in the detector is dominated by the instrumentation deployed in this sterile ice.

Each optical sensor consists of a glass sphere containing the photomultiplier and the electronics board that digitizes the signals locally using an onboard
computer. The digitized signals are given a global time stamp with residuals accurate to less than 3\,ns and are subsequently transmitted to the surface. 
Processors at the surface continuously collect the time-stamped signals from the optical modules, each of which functions independently. The digital messages are sent to a string processor and a global event builder. They are subsequently sorted into the Cherenkov patterns emitted by secondary muon tracks, or electron and
tau showers, that reveal the direction of the parent neutrino \cite{Halzen:2013bta}.

Based on data taken during construction, the actual neutrino-collecting area of the completed IceCube detector is larger by a factor 2 (3) at PeV (EeV) energy over what had been expected \cite{Ahrens2004}, mostly because of improvements to the data acquisition and analysis chain. The neutrino-collecting area is expected to increase further with improved calibration and development of optimized software tools for the detector, which has been operating stably in its final configuration since May 2011. Already reaching an angular resolution of better than 0.5 degree for muon tracks triggered, this resolution can be reduced off-line to $\leq 0.3$\,degree for individual events. The absolute pointing has been determined by measuring the shadowing of cosmic-ray muons by the moon to 0.1 degree at FWHM.

IceCube detects $10^{11}$ muons per year at a trigger rate of 2700\,Hz. Among these it filters $10^5$ neutrinos, one every six minutes, above a threshold of $\sim 100$\,GeV. The DeepCore infill array identifies a sample, roughly equal in number, with energies as low as 10 GeV; see Fig. \ref{deepcore}.  These muons and neutrinos are overwhelmingly of atmospheric origin. They are the decay products of pions and kaons produced by collisions of cosmic-ray particles with nitrogen and oxygen nuclei in the atmosphere. With larger detectors, the separation of cosmic-ray muons from secondary muons of neutrino origin becomes relatively straightforward even though their ratio is at the level of $10^6:1$. Muon tracks are reconstructed by likelihood methods and their energy deposition in the detector is measured in real time. High-purity neutrino samples of upgoing muon tracks of neutrino origin are separated from downgoing cosmic-ray muons by quality cuts; for instance, on the likelihood of the fit, on the number of photons that arrives at DOMs at the Cherenkov time (i.e., without a significant time delay resulting from scattering), on the length of the track, on the ``smoothness" requiring a uniform distribution of photoelectrons along the length of the track, etc. Each analysis produces appropriate cuts that depend on the magnitude of the background and the purity required to isolate a signal.

Atmospheric neutrinos are a background for cosmic neutrinos, at least at energies below 100 TeV. Above this energy, the flux is too small to produce events in a kilometer-scale detector; see Fig. \ref{discovery}. A small charm component is anticipated; its magnitude is uncertain and remains to be measured. As in conventional astronomy, IceCube must look through the atmosphere for cosmic neutrinos.

\section{Discovery of Cosmic Neutrinos}

The generation of underground neutrino detectors preceding construction of the AMANDA detector searched for cosmic neutrinos without success and established an upper limit on their flux, assuming an $E^{-2}$ energy dependence \cite{frejuslimit}:
\begin{equation}
E_\nu^2 \frac{dN}{dE_\nu} \leq 5\times10^{-9}\,\rm TeV\,cm^{-2}\,s^{-1}\,sr^{-1}
\label{eq:frejus}
\end{equation}
Operating for almost one decade, the AMANDA detector improved this limit by two orders of magnitudes. With data taken during its construction, IceCube's sensitivity rapidly approached the theoretical flux estimates for candidate sources of cosmic rays such as supernova remnants, gamma-ray bursts and, with a larger uncertainty, active galactic nuclei; see Fig.~\ref{discovery}. With its completion, IceCube also positioned itself for observing the much anticipated cosmogenic neutrinos with some estimates predicting as many as 2 events per year.

\begin{figure*}[t]
\begin{center}
\includegraphics[width=1.0\linewidth,trim=0px 70px 0px 40px,clip=true]{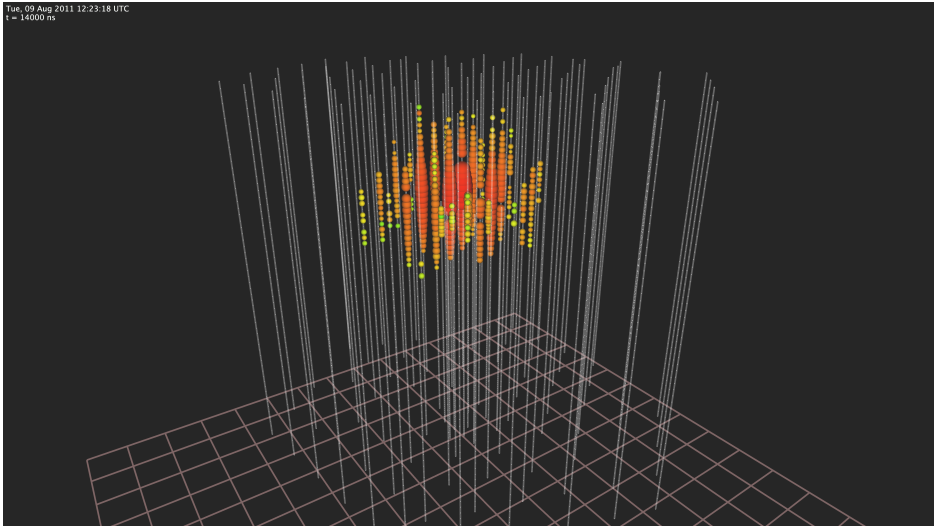}
\caption{Light pool produced in IceCube by a high-energy neutrino. The measured energy is 1.07 PeV, which represents a lower limit on the energy of the neutrino that initiated the shower. The vertical lines of white dots represent the sensors that report any detected signal. Color of the dots indicates arrival time, from red (early) to purple (late) following the rainbow. Size of the dots indicates the number of photons detected.}
\label{ernie}
\end{center}
\end{figure*}

Cosmogenic neutrinos were the target of a dedicated search using IceCube data collected between May 2010 and May 2012. Two events were found \cite{Aartsen:2013bka}. However, their energies, rather than super-EeV, as expected for cosmogenic neutrinos, were in the PeV range: 1,070 TeV and 1,240 TeV. These events are particle showers initiated by neutrinos interacting inside the instrumented detector volume. Their light pool of roughly one hundred thousand photoelectrons extends over more than 500 meters; see Fig. \ref{ernie}. With no evidence of a muon track, they are initiated by electron or tau neutrinos.

Previous to this serendipitous discovery, neutrino searches had almost exclusively specialized to the observation of muon neutrinos that interact primarily outside the detector to produce kilometer-long muon tracks passing through the instrumented volume. Although creating the opportunity to observe neutrinos interacting outside the detector, it is necessary to use the Earth as a filter to remove the huge background flux of muons produced by cosmic ray interactions in the atmosphere. This limits the neutrino view to a single flavor and half the sky. Inspired by the observation of the two PeV events, a filter was designed that exclusively identifies neutrinos interacting inside the detector. It divides the instrumented volume of ice into an outer veto shield and a 420 megaton inner fiducial volume. The separation between veto and signal regions was optimized to reduce the background of atmospheric muons and neutrinos to a handful of events per year while keeping 98\% of the signal. The great advantage of specializing to neutrinos interacting inside the instrumented volume of ice is that the detector functions as a total absorption calorimeter measuring energy with a 10--15\% resolution. Also, neutrinos from all directions in the sky can be identified, including both muon tracks produced in $\nu_\mu$ charged-current interactions and secondary showers produced by neutrinos of all flavors.

\begin{figure*}[t]
\begin{center}
\vspace{-25pt}
\includegraphics[width=1.0\linewidth,trim=0px 0px 0px -30px,clip=true]{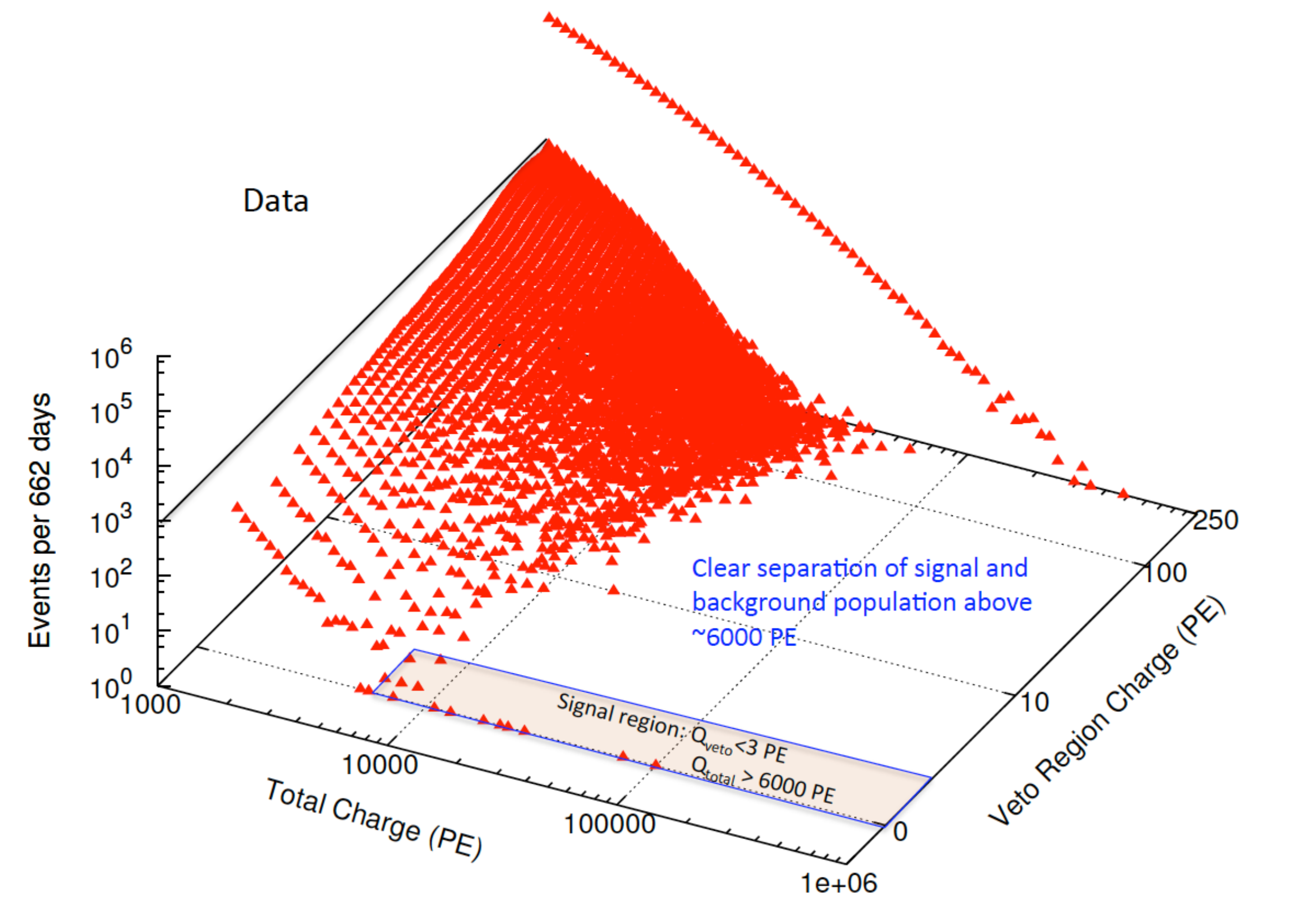}
\caption{Two years of IceCube data as a function of the total number of photoelectrons and the number present in the veto region. The signal region requires  more than 6000 photoelectrons with less than three of the first 250 in the veto region of the detector. The signal, including nine events with reconstructed neutrino energy in excess of 100\,TeV, is clearly separated from the background.}
\label{hesedata}
\end{center}
\end{figure*}

Analyzing the data covering the same time period as the cosmogenic neutrino search, 28 candidate neutrino events were identified with in-detector deposited energies between 30 and 1240 TeV; see Fig.\ref{hesedata}. Of these, 21 are showers whose energies are measured to better than 15\% but whose directions are determined to 10-15 degrees only. Predominantly originating in the Southern Hemisphere, none show evidence for a muon track. If atmospheric in origin, the neutrinos should be accompanied by muons produced in the air shower in which they originate. For example, at 1 PeV, less than 0.1\% of atmospheric showers contain no muons with energy above 500 GeV, approximately that which is needed to reach the detector in the deep ice when traveling vertically.

The remaining seven events are muon tracks, which do allow for subdegree angular reconstruction; however, only a lower limit on their energy can be established because of the unknown fraction carried away by the exiting muon track. Furthermore, with the present statistics, these are difficult to separate from the competing atmospheric background. The 28 events include the two PeV events previously revealed in the cosmogenic neutrino search. The signal of 28 events on an atmospheric background of $10.6^{+5.0}_{-3.6}$ represents an excess over background of more than 4 standard deviations.

\begin{figure*}[htbp]
\begin{center}
\vspace{-25pt}
\includegraphics[width=1.0\textwidth]{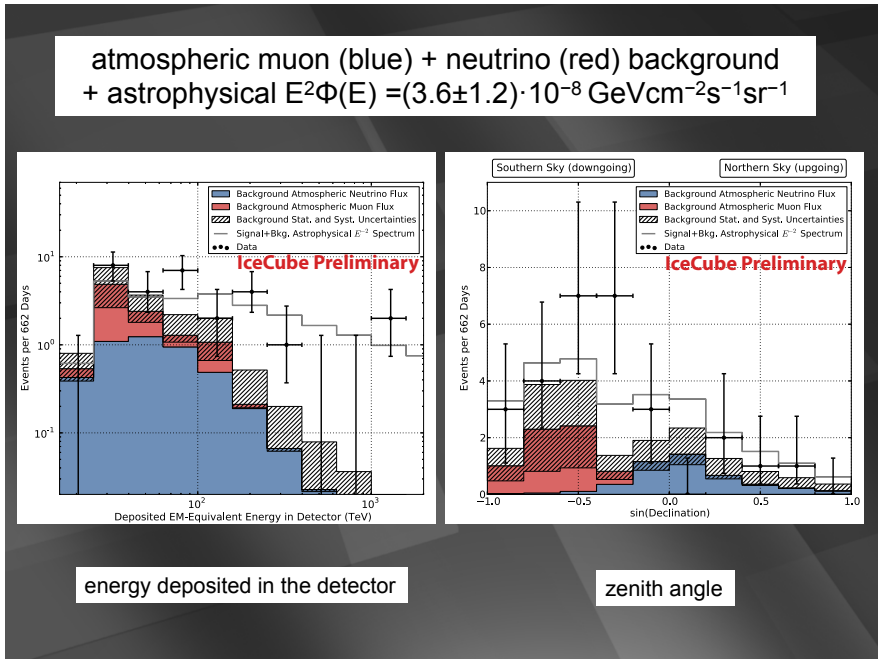}
\caption{Distribution of the deposited energies (left) and declination angles (right) of the observed events compared to model predictions. Energies plotted are in-detector visible energies, which are lower limits on the neutrino energy. Note that deposited energy spectra are always harder than the spectrum of the neutrinos that produced them due to the neutrino cross section increasing with energy. The expected rate of atmospheric neutrinos is based on Northern Hemisphere muon neutrino observations. The estimated distribution of the background from atmospheric muons is shown in red. Due to lack of statistics from data far above our cut threshold, the shape of the distributions from muons in this figure has been determined using Monte Carlo simulations with total rate normalized to the estimate obtained from our in-data control sample. Combined statistical and systematic uncertainties on the sum of backgrounds are indicated with a hatched area. The gray line shows the best-fit $E^{-2}$ astrophysical spectrum with all-flavor normalization (1:1:1) of $E_\nu^2 \frac{dN}{dE_\nu}=3.6\times10^{-11}\,\rm TeV\,cm^{-2}\,s^{-1}\,sr^{-1}$ and a spectral cutoff of 2 PeV.}
\label{hese_energy_zenith}
\end{center}
\end{figure*}

The large errors on the background are associated with the possible presence of a neutrino component originating from the production and prompt leptonic decays of charmed particles in the atmosphere. Such a flux has not been observed so far. While its energy and zenith angle dependence are known, its normalization is not; see Fig. \ref{discovery} for one attempt at calculating the flux of charm origin. Neither the energy, nor the zenith angle dependence of the 28 events observed can be described by a charm flux, and, in any case, fewer than 3.4 events are allowed at the 1\,$\sigma$ level by the present upper limit on a charm component of the atmospheric flux set by IceCube itself \cite{schukraft}.  As already mentioned, in the case of a charm origin, the excess events should contain accompanying muons from the atmospheric shower that produced them, but they do not. Fitting the data to a superposition of extraterrestrial neutrinos on an atmospheric background yields a cosmic neutrino flux of

\begin{equation}
E_\nu^2 \frac{dN}{dE_\nu}=3.6\times10^{-11}\,\rm TeV\,cm^{-2}\,s^{-1}\,sr^{-1}
\label{eq:heseflux}
\end{equation}
for the sum of the three neutrino flavors. As discussed in section 2, this is the level of flux anticipated for neutrinos accompanying the observed cosmic rays. Also, the energy and zenith angle dependence observed is consistent with what is expected for a flux of neutrinos produced by cosmic accelerators; see Fig. \ref{hese_energy_zenith}. The flavor composition of the flux is, after corrections for the acceptances of the detector to the different flavors, consistent with 1:1:1 as anticipated for a flux originating in cosmic sources.

\begin{figure*}
\begin{center}
\vspace{-25pt}
\includegraphics[width=1.0\linewidth,trim=0px 40px 0px -10px,clip=true]{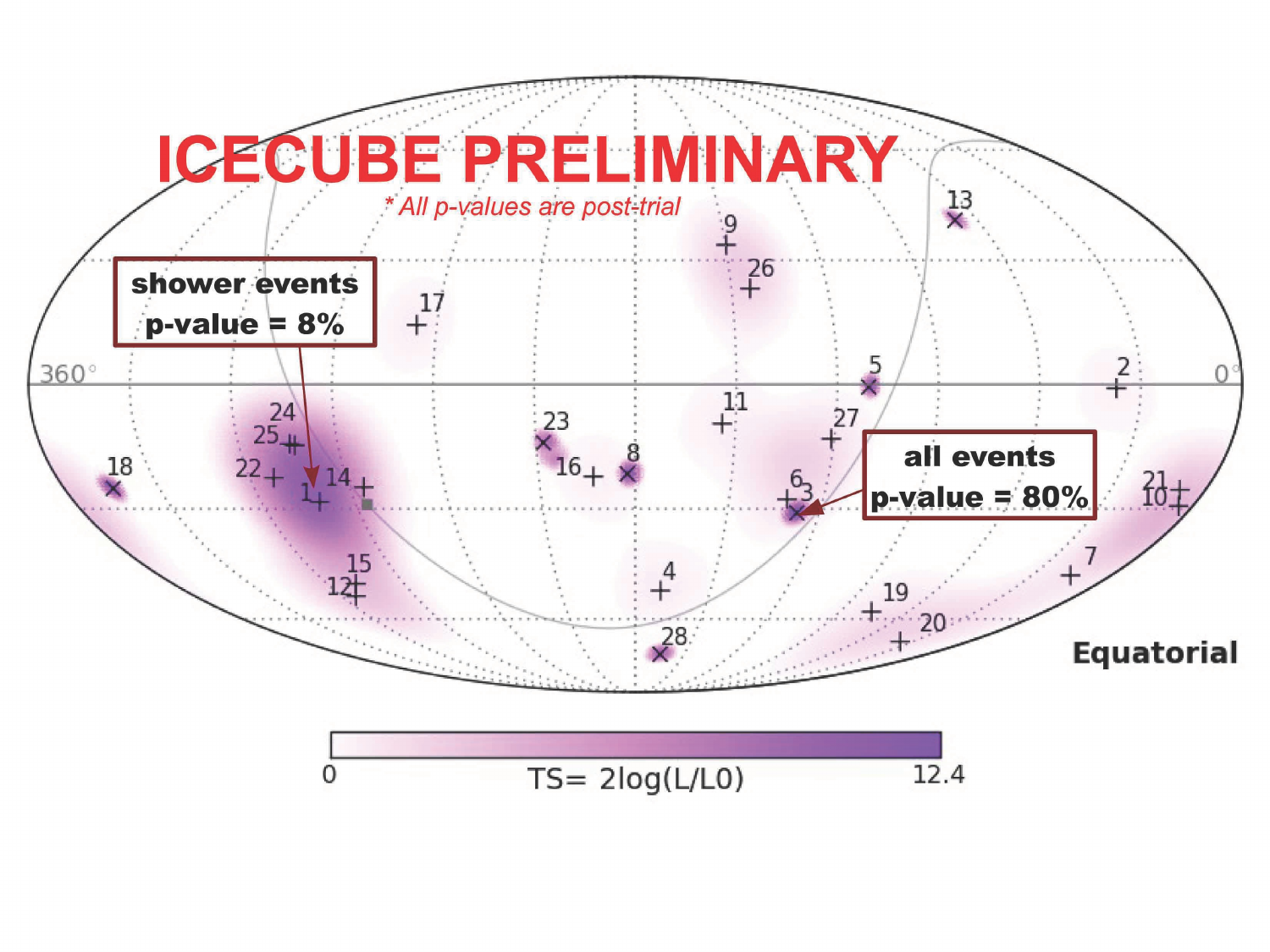}
\caption{Sky map in equatorial coordinates of the test statistic (TS) that measures the probability of clustering among the 28 events. The most significant cluster consists of five events---all showers and including the second-highest energy event in the sample---with a final significance of only 8\%. The Galactic plane is shown as a gray line with the Galactic center denoted as a filled gray square. Best-fit locations of individual events are indicated with vertical crosses (+) for showers and angled crosses (x) for muon tracks.}
\label{hesemap}
\end{center}
\end{figure*}

So, where do the neutrinos come from? A map of their arrival directions is shown in Fig. \ref{hesemap}. We used a test statistic $TS=2 \times log{L/L_0}$, where L is the signal plus background maximized likelihood and $L_0$ is the background only likelihood obtained by scrambling the data. No significant spot on the sky was found when compared to the randomized pseudo experiments. Repeating the analysis for showers only, a hot spot appears at RA=281 degrees and dec=23 degrees close to the Galactic center. After correcting for trials, the probability corresponding to its TS is 8\%. We also searched for clustering of the events in time and investigated a possible correlation with the times of observed GRBs. No statistically significant correlation was found. Fortunately, more data is already available, and the analysis, performed blind, can be optimized for searches of future data samples.

For additional information, see \cite{kopper}.

\section{Conclusions: Too Early to Speculate?}

That the present information is insufficient to identify the sources of these events is illustrated by the range of speculations in the literature \cite{Lipari:2013taa,Bhattacharya:2011qu,Laha:2013lka,Kistler:2013my,Anchordoqui:2013lna,He:2013zpa,Winter:2013cla,Roulet:2012rv,Kalashev:2013vba,Murase:2013ffa,Stecker:2013fxa,Murase:2013rfa,He:2013cqa,Fox:2013oza,Neronov:2013lza,Feldstein:2013kka,Esmaili:2013gha}. The first question to answer is whether the excess neutrinos are Galactic or extragalactic in origin.

If the observed flux is truly diffuse, it is most likely of extragalactic origin. While the statistics are not compelling, the excess events seem to originate mostly in the Southern Hemisphere. Furthermore, seven shower events cluster in one of the $30^\circ \times 30^\circ$ bins shown on the map in Fig. \ref{hesemap}, where 0.6 are expected for a uniform distribution. While the cluster seems to be displaced from the Galactic center, the second highest-energy event of 1.07\,PeV does reconstruct in this direction. 

If cosmic accelerators are the origin of the excess flux, then the neutrinos have been produced in proton-photon or proton-proton interactions with radiation or gas at the acceleration site or along the path traveled by cosmic rays to Earth. The fraction of energy transferred to pions is about 20\% and 50\% for $p\gamma$ and $pp$, respectively, and each of the three neutrinos from the decay chain $\pi^+\to\mu^+\nu_\mu$ and $\mu^+\to e^+\nu_e\bar\nu_\mu$ carries about one quarter of the pion energy. Hence, the cosmic rays producing the excess neutrinos have energies of tens of PeV, well above the knee in the spectrum. It is tantalizingly close to the energy of $100$~PeV~\cite{Apel:2013dga,IceCube:2013wda} where the spectrum displays a rich structure, sometimes referred to as the ``iron knee." While these cosmic rays are commonly categorized as Galactic, with the transition to the extragalactic population at the ankle in the spectrum at  $3 \sim 4$\,EeV, one cannot rule out a subdominant contribution of PeV neutrinos of extragalactic origin. IceCube neutrinos may give us information on the much-debated transition energy.

The flux observed by IceCube is close to the Waxman-Bahcall bound \cite{Waxman:1998yy,Bahcall:1999yr}, which applies to extragalactic sources transparent to photons. For these, the energy in cosmic rays translates into an upper limit on the neutrino flux. To accommodate the observed cosmic rays above the ankle, the accelerators must generate an energy of $\simeq 2 \times 10^{37}\,\rm\,erg\ s^{-1}$ per Mpc$^3$, as previously discussed. This translates into a 3-flavor neutrino flux:
\begin{equation}
E_\nu^2 \frac{dN}{dE_\nu}=2 (5) \times10^{-11} \times \xi_z\ \,\rm TeV\,cm^{-2}\,s^{-1}\,sr^{-1},
\label{eq:wb}
\end{equation}
for $p\gamma$ ($pp$) neutrino-producing interactions. The factor $\xi_z \simeq 3$ takes into account the evolution of the sources as a function of their redshift. The IceCube excess saturates this bound, at least for hadronic origin of the neutrinos. However, if the neutrinos are indeed produced by 100 PeV cosmic rays, the lower transition energy to extragalactic cosmic rays results in a larger energy requirement for the production of the extragalactic cosmic rays and an increase of the bound by an order of magnitude \cite{Waxman:1998yy}. This leaves the IceCube flux as a subdominant component; see, however,~\cite{Katz:2013ooa}.

Whether of $pp$ and $p\gamma$ origin, neutrinos are accompanied by $\gamma$-rays that are the decay products of neutral pions produced in association with the charged ones. While no TeV--PeV gamma rays of pionic origin have been observed so far, experiments have established limits on a possible PeV gamma-ray flux, independent of its origin \cite{Schatz:2003aw,Matthews:1990zp,Chantell:1997gs,Aartsen:2012gka}.

The relative flux of neutrinos and $\gamma$-rays is determined by the ratio of charged to neutral pion secondaries, $K$.  In the case of $pp$ interactions  $K\simeq2$ while for $p\gamma$ interactions the number of $\pi^+$ and $\pi^0$ secondaries is roughly equal, hence $K\simeq1$. For a transparent source, we have 
\begin{equation}\label{eq:Jrel}
E_\gamma \frac{dN_\gamma}{dE_\gamma}(E_\gamma) \simeq e^{-\frac{d}{\lambda_{\gamma\gamma}}}\frac{2}{K} \frac{1}{3} \,E_\nu \frac{dN_\nu}{dE_\nu}(E_\nu)\,,
\end{equation}
where the neutrino flux is the all-flavor flux of Eq. \ref{eq:heseflux}. The proportionality factor between the fluxes is K; the other factors in the above relation correct for the fact that: i) TeV--PeV $\gamma$-rays, unlike neutrinos, are absorbed in radiation backgrounds with interaction length $\lambda_{\gamma\gamma}(E_\gamma)$, ii) in the decay $\pi^0\to\gamma\gamma$ the $\gamma$-ray takes half of the pion energy, and  iii) each of the three neutrinos from charged pion decay carries about one quarter of the pion's energy. The gamma-ray flux accompanying Eq. \ref{eq:heseflux} is in conflict with the upper limits of the CASA-MIA \cite{Matthews:1990zp,Chantell:1997gs} and KASCADE \cite{Schatz:2003aw} experiments. Because PeV photons reach us from Galactic distances as large as 10\,kpc with modest attenuation, this apparently excludes the Galactic origin of the IceCube flux \cite{ahlersmurase}. This conclusion does, however, depend on the assumption that the sources are transparent to gamma rays and that the flux is isotropic. Specifically, Galactic origin cannot be ruled out for subclasses of events like the cluster of events near the Galactic center previously mentioned. Finally, all experiments besides IceCube are in the Northern Hemisphere, leaving a blind spot in the sky that contains more than half of the IceCube events. For an in-depth discussion of the gamma-ray limits, see \cite{ahlersmurase}.

If, in fact, any of the IceCube events observed in the blind spot do originate from a Galactic point source, IceCube itself should be able to observe the accompanying PeV gamma rays. These are detected as muon-poor showers triggered by IceTop. The level of point-source flux per neutrino flavor corresponding to one out of the 28 events is given by
\begin{eqnarray}
E_\nu \frac{dN}{dE_\nu}= 4\pi \, \frac{1}{28} \,1.2\times10^{-11}\,\rm cm^{-2}\,s^{-1}  \nonumber \\
\simeq 5.4\times10^{-12}\,\rm cm^{-2}\,s^{-1},
\label{eq:psflux}
\end{eqnarray}
with the corresponding pionic photon flux a factor of 2 larger, assuming $pp$ interactions; see above. This is a flux of $\sim10^{-17}\,\rm cm^{-2}\,s^{-1}$ at 1\,PeV, well within the gamma-ray sensitivity of the completed IceCube detector; see Fig. 15 in \cite{Aartsen:2012gka}. In fact, the highest fluctuation in a gamma-ray map obtained with one year of data collected with the detector when it was half complete is in the direction of one of the PeV neutrino events \cite{ahlersmurase}.

The acceptance of the starting event analysis producing the first evidence for an extraterrestrial neutrino flux means the signal consists 
mostly of electron and tau neutrinos originating in the Southern Hemisphere. In contrast, a detector in the Mediterranean views the Southern Hemisphere through the Earth and therefore has sensitivity to muon neutrinos that can be reconstructed with subdegree precision. For illustration~\cite{viviana}, an IceCube detector cloned and positioned in the Mediterranean would observe 71 muon neutrinos per year with energy in excess of 45\,TeV, for a muon neutrino flux of
\begin{eqnarray}
E^2_\nu \frac{dN_{\nu_\mu + \bar{\nu}_\mu}}{dE_\nu} = 1.2 \times 10^{-11} \quad {\rm TeV}~{\rm cm}^{-2}~{\rm s}^{-1}~{\rm sr}^{-1}\,.
\end{eqnarray}
We here assume a 1:1:1 distribution of flavors, which is consistent with observation. Above 45\,TeV, signal should dominate, providing a sky map with little background. The event rate is likely to be an overestimate because the effective area for a diffuse analysis, which typically requires stronger cuts on the data, is smaller than the point source area used here. In the case of IceCube this correction is close to a factor of two.

If the cluster of seven events close to the center of the Galaxy, referred to above, originated from a point source, the corresponding flux would be 
\begin{eqnarray}
E^2_\nu \frac{dN_{\nu_\mu + \bar{\nu}_\mu}}{dE_\nu} = 6 \times 10^{-11}\quad {\rm TeV}~{\rm cm}^{-2}~{\rm s}^{-1},
\end{eqnarray}
yielding 45 events per year. This flux is simply estimated by multiplying the diffuse flux by $4\pi \times 7/ [28-10.6]$, where we corrected for the number of background events events in the sample of 28. The number is not corrected for the fact that the center of the Galaxy is only visible 68\% of the time for a Mediterranean detector.

Both the diffuse and point source signals would be statistically significant within one year. The operating Antares detector is a factor of 40 smaller than the IceCube detector, and therefore the IceCube excess only produces signals at the one-event level per year. For a possible point source associated with 7 events, the predicted flux is actually close to the present IceCube and Antares limits towards the center of the Galaxy. Larger event samples, especially of well-reconstructed muon neutrinos, are likely to be the key to a conclusive identification of the origin of the IceCube extraterrestrial flux. If the flux observed in IceCube turns out to be isotropic, IceCube itself will observe the same diffuse muon neutrino signal from the Northern Hemisphere.

In summary, it may be too early to speculate. IceCube already collected one more year of data and was designed to operate for 20 years. The analysis, done blind, can now be optimized on the signal observed. Several such analyses, including several optimized for muon neutrinos are already underway.

\section{Acknowledgements}
Discussion with collaborators inside and outside the IceCube Collaboration, too many to be listed, have greatly shaped this presentation. Thanks.
This research was supported in part by the U.S. National Science Foundation
under Grants No.~OPP-0236449 and PHY-0969061and by the University of Wisconsin Research
Committee with funds granted by the Wisconsin Alumni Research Foundation.



%
\end{document}